\documentclass[9pt,shortpaper,twoside,web]{ieeecolor}
\usepackage{apg}
\makeatletter\let\NAT@parse\undefined\makeatother %correct the fake natbib commands hack found
\usepackage{amsmath,amssymb,amsfonts}
\usepackage{algorithmic}
\usepackage{graphicx}
\usepackage{textcomp}
\usepackage[switch]{lineno}

\def\BibTeX{{\rm B\kern-.05em{\sc i\kern-.025em b}\kern-.08em
    T\kern-.1667em\lower.7ex\hbox{E}\kern-.125emX}}
\usepackage[authoryear]{natbib} % Load natbib with options for author-year
\usepackage{cite}
\usepackage{booktabs,multirow,accents}

\usepackage{bm}
  
\usepackage{amsthm}

\theoremstyle{definition}

\theoremstyle{remark}

\makeatletter\def\operator@font{\sf}\makeatother

\usepackage{lastpage,fancyhdr}
\usepackage{ifthen,hyperref}
\usepackage{lipsum}
\fancyheadoffset[L]{0.25in}
\fancyfootoffset[L]{0.25in}
\title{PowerLine Unmanned Surfer (PLUS):\\Shape Adaptive Dynamics Development and Control Design}
\author{Ujjval Patel, Francis Phillips, Todd Henry, and Imraan Faruque.
\thanks{U. N.~Patel, and I. A. Faruque are with Oklahoma State University, Stillwater, OK 74078 USA (e-mail: ujjval@okstate.edu, i.faruque@okstate.edu). }
\thanks{F. R. Phillips, and T. C. Henry are with DEVCOM Army Research Laboratory, 800 Park Offices Drive, Durham, NC  27709 USA (e-mail: francis.r.phillips7.civ@army.mil ,todd.c.henry2.civ@army.mil).}
\thanks{The Army Research Lab partially supported this work under ARL-R-WMRD-300124. Copyright by the authors May 2024; this work may be under consideration for publication and copyright may be transferred without notice, after which this version may no longer be accessible.}}

\begin{document}
\maketitle
\begin{abstract}This paper introduces the powerline unmanned surfer (PLUS) concept to extend the limited endurance of fixed wing unmanned aerial vehicles (UAVs) via in-flight energy harvesting from overhead electrical distribution power lines, and develops the flight dynamics and control framework to support centimeter-scale longitudinal powerline frequency tracking.  The dynamics framework models the UAV's shape adaptive structure, aerodynamic forces, and control inputs, and applies the coupled flight mechanics framework to low clearance tracking of powerline contours. This study develops a ``trajectory to shape-adaptive UAV'' controller design approach for longitudinal powerline tracking through local spatial frequency matching. The frequency-matching approach dynamically regulates the aircraft modes' frequency to the powerline catenary spatial frequency using a generalized parameter linearization approach. Performance is assessed on an example UAV implementing camber and thickness morphing by quantifying clearance distance from neighborhood to high voltage powerline environments and across span and chord combinations. This approach achieves alternating periods of low-clearance tracking and antiphase oscillation, with 34\% of the powerline having tracking error less than 1m. Airfoil thickness increase has a stronger effect over Phugoid mode wavelength than thickness decrease. Wingspan, chord, and powerline parameter sensitivities are quantified, providing a foundation for near-field long-distance powerline tracking and energy harvesting.\end{abstract}
\section{Introduction}

Electric propulsion unmanned aerial vehicles (UAVs) have become integral in applications such as surveillance and payload delivery, yet their endurance and range are limited by the energy constraints of batteries. Although conventional aerial refueling offers partial solutions, its limited efficiency and infrastructure requirements highlight the need for innovative approaches.  
%\twocolumn
Traditional UAV design approaches, particularly for fixed wings, often begin with a set of aircraft performance requirements, such as range, endurance, and cruise speed, and later develop maneuverability limits and achievable trajectories via turn rate and speed limitations.  This approach focuses on gross aircraft performance rather than precisely tailoring the airframe's dynamic maneuverability characteristics to a specific mission trajectory.  Fundamentally, the dynamically achievable trajectories are not precisely specified as input and are an iteration or an envelope specified by its corners rather than the natural behaviors of the aircraft.

As UAV missions become more diverse, including new prescribed flight maneuvers and trajectories, mechanisms to incorporate maneuverability and trajectory-following in early stage design will be helpful.  Applications could include terrain following, flight in caves, or flight in cluttered environments.  In such examples, instead of "what are the achievable trajectories for a given configuration?" the design for extreme behaviors asks "what airframe configuration(s) are able to achieve a given trajectory?" In some cases, the integration of an additional effector, such as a morphing structure or actuator, could provide a required envelope expansion.

This design approach provides an alternative inverse to traditional configuration design, in which an airframe's detailed maneuverability characteristics is an outcome of configuration choice and sizing. The design-for-trajectory approach instead uses mission trajectory specification as an input to specify the airframe's outer mold line, which may include new shape adaptation modalities to achieve new trajectories.

This study introduces the powerline unmanned surfer (PLUS) concept as an early example of this dynamics-integrated UAV design, beginning with mission-specific trajectory requirements. The focus of PLUS on powerline tracking supports UAV endurance improvements by enabling in-flight energy harvesting from overhead electric powerlines. An integrated approach that encompasses structure, aerodynamics, and controls allows prolonged flight and improved mission performance.

A core challenge in the implementation of PLUS lies in the control necessary to track the discontinuous trajectories of the powerlines while maintaining electrical connectivity. This demands an integrated control and dynamics framework, balancing the UAV's adaptive structure, aerodynamic forces, and control input. The focus of this paper is on this dynamic development aspect-- a framework to incorporate airframe flight dynamics and with shape adaptation and assess trajectory feasibility. % the development of shape-adaptive airframes for precise trajectory tracking along powerlines, ensuring continuous energy harvesting and extended flight durations.

This dynamics framework integrates shape adaptive features into a reference airframe, aiming to explore the benefits and challenges of shape adaptiveness in altitude-based powerline tracking. Camber and thickness variation examples are used to explore the relationship between shape adaptiveness and natural aircraft longitudinal flight modes: Phugoid and short period.  The natural flight modes of the aircraft are then dynamically tailored to match the local spatial frequency of the hanging utility lines.

%The goal is a UAV capable of autonomously navigating and maneuvering around powerlines, using its dynamically shaped structures to maintain consistent energy harvesting.

A novel trajectory-to-shape adaptive design framework for longitudinal powerline tracking has been introduced, overcoming the limitations of conventional airframes in matching the catenary spatial frequency of powerlines. This framework represents a significant advance in UAV design, enabling real-time morphing of the airfoil geometry to meet the specific demands of powerline tracking. The open-loop control approach is applied to a shape adaptive air vehicle simulation to quantify its performance across a range of utility line environments.

\subsection{Previous Work}
The PLUS concept connects several areas in which UAVs have received considerable research and development, particularly in the areas of endurance enhancement, adaptive structures, and powerline monitoring. 

\subsubsection{UAV Endurance Enhancement}
Structural adaptation to increase the control authority of fixed wing small UAV is an idea often modeled by biological systems \citep{HarveyNature, HarveyPAS, CheneyPRSL, QuinnPNAS}. 
Considerable previous attention \citep{seigler2009transitionStabilityMorphing,kota2003compliantMorphingDesign} including several previous surveys \citep{barbarino2011surveyMorphing,seigler2005morphingDynCtrl, jha2004surveyClassificationMorphing}. Adding to these detailed surveys is outside the scope of this paper.  

Instead, we note that the range of engineered adaptation (and consequently the resulting trajectory) is typically less aggressive than the shape adaptations seen in avian systems. For example, \citep{CheneyPRSL} used small wing twist and camber adjustments complemented by tail movements to maintain an optimal trajectory. Compared to conventional UAV, avian systems have higher degrees of articulation using smooth, active surfaces that lack control linkages and hinges at the outer mold line. The degree to which smooth active surfaces can be used to replace various internal/external control surfaces in the wing and tail has been researched by \citep{Pankonien,Bilgen2013}.  Even at the comparatively conservative trajectories seen in previous engineered work, these adaptations have suggested a need for a new design philosophy \citep{ajaj2016morphingNeedsNewDesignPhilosophy}. As UAV missions with more demanding trajectories are addressed as done by \citep{cohen1} and \citep{cohen2}, the need for a detailed ``trajectory-to-aerodynamic-shape'' specification method becomes a more significant limitation.

\subsubsection{Powerline Monitoring Technologies}
Powerline monitoring has been a critical application area for UAVs. \citep{quad2} explored the use of multirotor UAVs for close-range powerline inspections, emphasizing the importance of stable flight dynamics near high-tension lines. The integration of advanced event based sensors for fault detection in powerline infrastructure was highlighted in \citep{quad1}, underscoring the utility of UAV in preventative maintenance.

Flight in close proximity to powerlines poses specialized control challenges. Adaptive control algorithms developed in \citep{quad3} for UAVs operating in complex environments laid the foundation for precision flight controls.  However, these studies focused on the use of quadcopters for powerline inspection and survey while maintaining a safe distance from the powerline infrastructure. These studies did not seek to leverage powerline structures as a means of enhancing UAS capabilities. 

\subsubsection{Powerline energy harvesting}
In-flight powerline energy harvesting could be accomplished from a moving fixed wing UAV via either a trailing contact (``shoe'') or an inductive coil. The magnetic field of a long straight wire is 
\[B=\frac{\mu_0 I}{2\pi R},\] where $I$ is the current, $R$ the distance to the wire, and $\mu_0=4\pi\times10^{-7}\text{T m/s}$ the permeability of free space.  A single 345kV powerline carrying 628A could have a 125$\mu$T magnetic field at 1m \citep{ruralUsdaEmFields}.
Inductive powerline energy harvesting has previously been approached for wireless sensor networks, using air-core coil and resonant capacitor circuits to demonstrate 1-6mW energy harvesting from a 21$\mu$T magnetic field switching at 60Hz \citep{tashiro2011powerlineEnergyHarvesting}.  These studies showed that the number of coil turns and core material affect the harvesting potential and estimated that energy harvesting levels at publicly acceptable levels could be comparable to photovoltaic cell output on a cloudy day ($130\mu$W/cm\textsuperscript{3} at 200$\mu$T) \citep{tashiro2011powerlineEnergyHarvesting,wang2021energyHarvestingCircuit}. 
Previous work tailored towards wireless sensor networks has shown scavenging at single amp levels, scavenging via inductors between parallel wires (850$\mu$W from 8.4A differential currents), and highlighted the importance of core choice for higher power densities (100mW/cm$^3$) \citep{riba2022powerlineHarvestingReview}.

A UAV in flight may not enclose the wire completely, requiring the inclusion of a demagnetization factor. 
 Work on non-enclosing energy harvesters has previously computed shape optimizations for such cases, such as the bow-tie shaped non-enclosing coil in \citep{yuan2015bowtiePowerlineHarvester}.

Research efforts have explored various approaches to enabling UAVs to interact with power lines for energy harvesting and recharging purposes, focusing on the development of mechanisms for physically grasping power lines. \citep{grapple1} focused on the design and testing of a mechanical grappling mechanism that allows UAVs to secure their connection to powerlines, providing a stable platform for energy transfer. \citep{grapple2} evaluated the feasibility of a drone charging function on overhead power lines and verified this in a test setup. The work of \citep{grapple3} introduced a novel mechanical structure attached to the nose of a light-weight fixed wing aircarft that could extract electrical energy from high-voltage powerlines with minimal loss and convert it to a form that could be used by UAV batteries. \citep{grapple4}’s presented a novel electromechanical recharging station that can be mounted on energized AC power line to charge the drone battery wirelessly without the need to modify the electrical infrastructure. Meanwhile, \citep{grapple5} explored the long-term impacts of continuous powerline energy harvesting on UAV battery life and performance, offering insights into optimal charging cycles and energy management strategies. Finally, \citep{grapple6} presented technologies that allow UAVs to autonomously attach and recharge from the existing DC railway infrastructure.

\subsubsection{Adaptive Structures for Maneuver}
Fixed wing UAVs typically achieve higher range and endurance relative to similar scale multirotor configurations that can achieve more decoupled motions \citep{Mulgaonkar2014QuadrotorPower}. One approach to mission endurance improvement  is perching at a speed that biological systems typically do with precise control of their velocity and position \citep{KleinH}. \citep{KleinH} suggested that trajectory information in combination with aeromechanical information (pressure/strain) \citep{Castano, Haughn} and perception (distance/location) could be used to optimize behavior in real time for UAV. \citep{Moore} showed that at speed perching could be done with conventional control surfaces with \citep{Greatwood}, and \citep{Fletcher} using sweep control and reinforcement learning to command the desired trajectory. Both \citep{Greatwood} and \citep{Fletcher} were able to, in real-world flight tests, command an approach and flare trajectory near a perch position with reasonable accuracy using reinforcement learning. \citep{Ajanic} developed a vehicle capable of aggressive maneuvering by creating a wing and tail capable of adaptation. 

The PLUS concept distinguishes itself from previous research studies by taking advantage of existing powerline catenary shape, thereby extending the existing capabilities of a fixed-wing UAV. %In addition, this paper goes into detail to simplify the design process.

\section{Methods \& Approach}
\subsection{Mission space definition}
\begin{figure*}[h!]   \centering
    \includegraphics[width=0.75\textwidth]{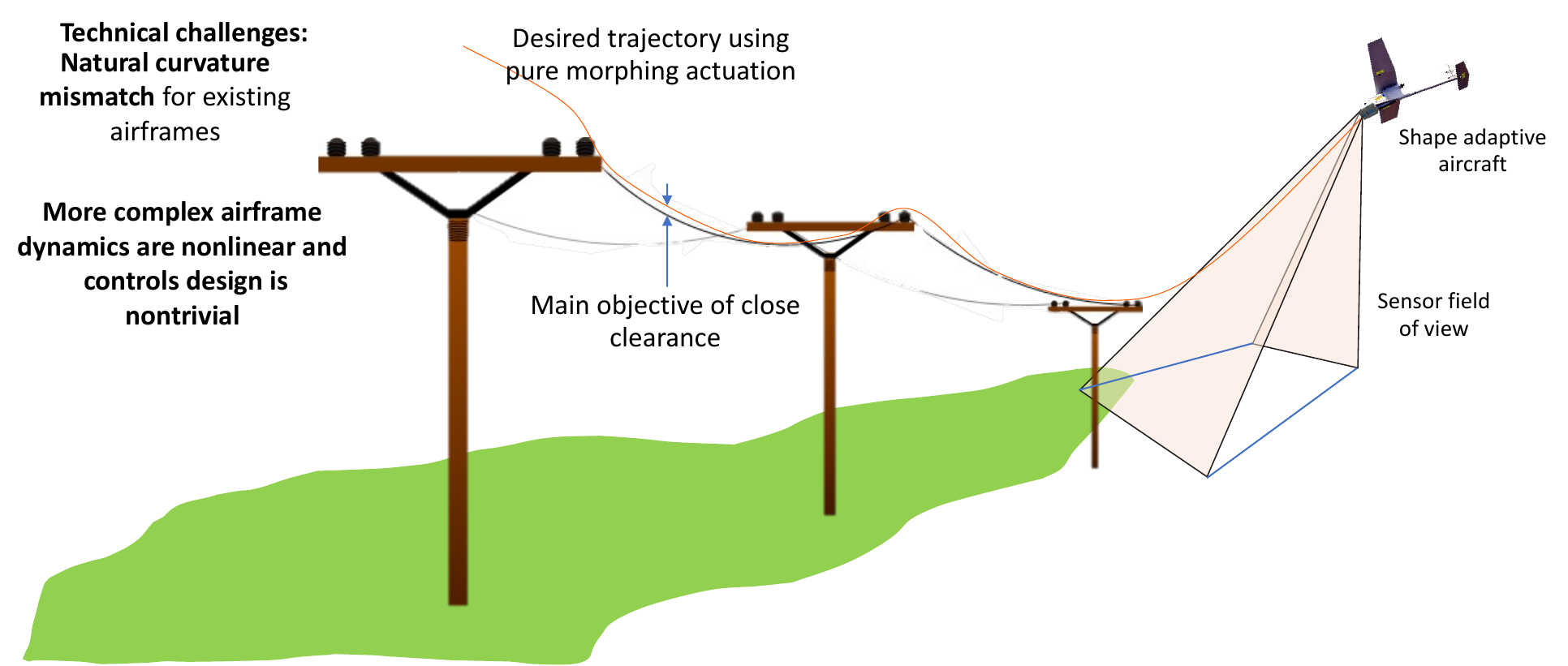}
    \caption{Powerline unmanned surfing concept for long range fixed wing UAV range extension and navigation.}  \label{fig:missionspace}\end{figure*}

\begin{figure}[!hbtp] \centering
    \includegraphics[width=0.45\textwidth]{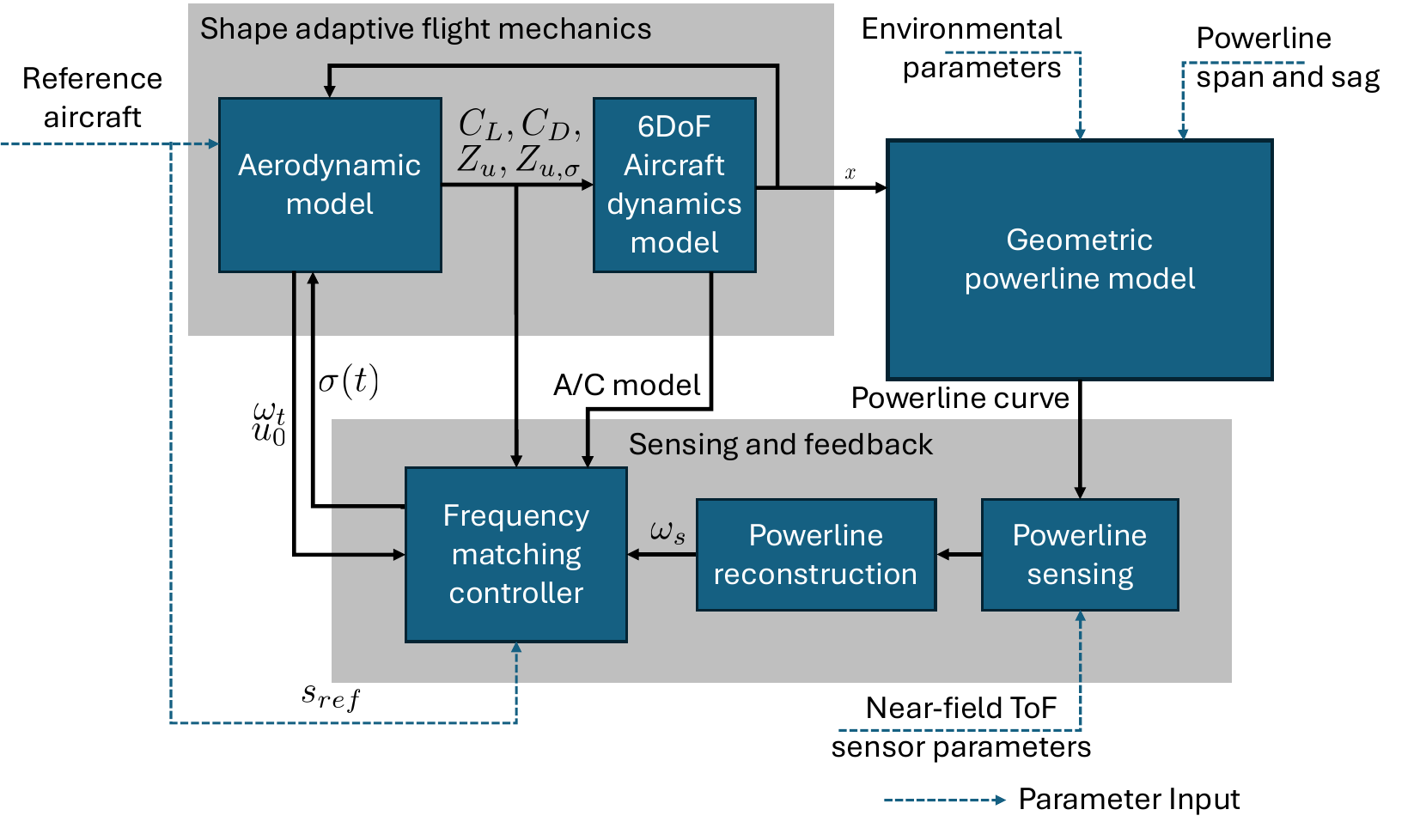}
    \caption{This paper's coverage: flight dynamics model development, powerline geometry environment, and controller development.}    \label{fig:PLUS}\end{figure}

A central feature of this framework is its precision in tracking the catenary contour of powerlines through "local spatial frequency matching." This process entails the agile adaptation of the UAV's spatial frequencies to align with the undulating spatial frequency of the powerlines. The research details a comprehensive dynamic model, accompanied by simulations to quantify the UAV's longitudinal tracking performance along powerline trajectories. Figure~\ref{fig:missionspace} illustrates the operational mission space of the PLUS system, and highlights technical challenges to the PLUS concept. 
Figure~\ref{fig:PLUS} illustrates the coverage of this paper, showing the interconnection between the flight dynamics model, powerline geometry, and sensing and feedback structure.

\subsection{Powerline Modeling}
\begin{figure}[!hbtp]\centering
    \includegraphics[width=0.45\textwidth]{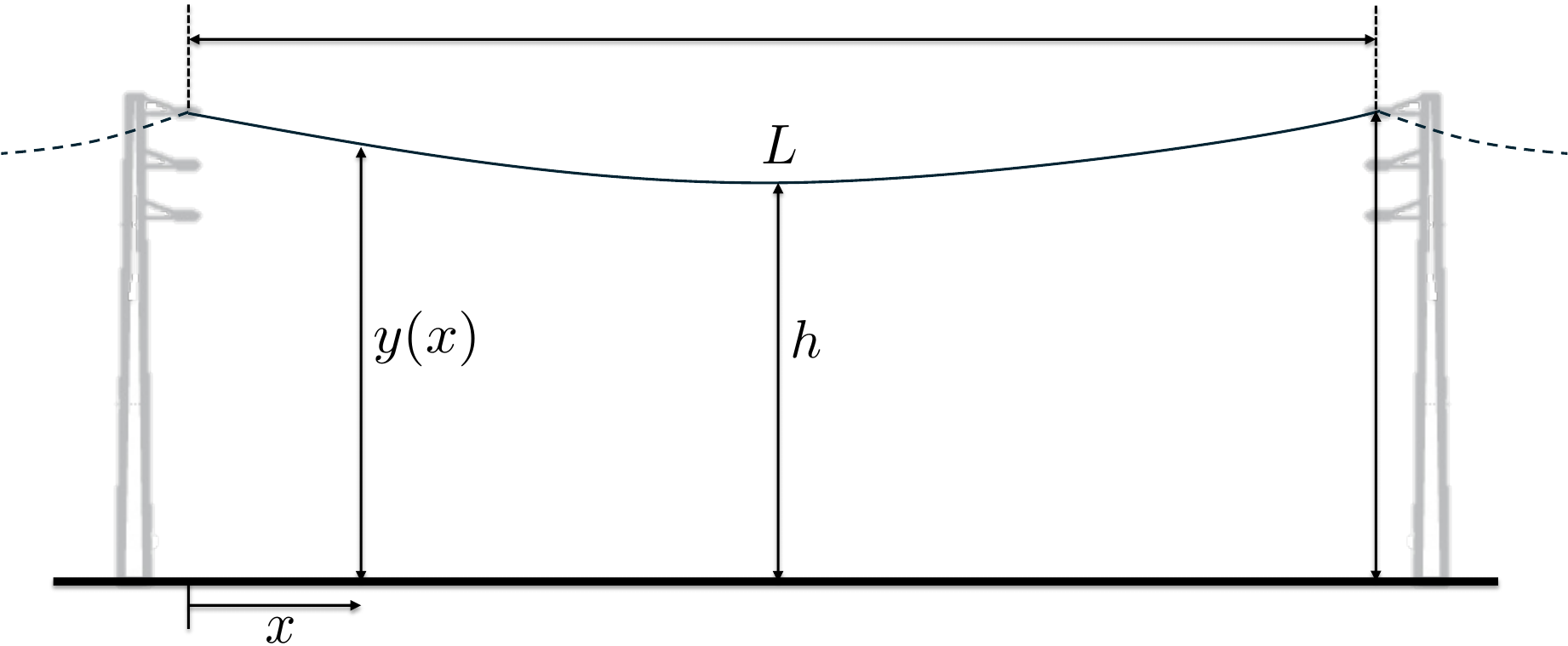}
    \caption{Powerline geometry used in this study \citep{tranline_designmanual}.}
    \label{fig:catenary}\end{figure}

An example powerline catenary curve is illustrated in Fig.~\ref{fig:catenary}, in which the between-pole curves are modeled as a catenary influenced by environmental parameters \citep{tranline_designmanual}.

The expression for a powerline catenary curve was given by
\begin{equation}\label{eq:catenary}
y = \dfrac{T_0}{\rho g A} \cosh{\left\{ \dfrac{\rho g A}{T_0} \left(x - \dfrac{L}{2}\right)\right\}} \quad \text{for } x \in \left[0,L\right],\end{equation}
where $T_0$ is the tension in the cable at ambient temperature, $\rho$ is the density of material in the transmission cable, $A$ is the cross-sectional area of the transmission cable and $g$ is gravitational acceleration.  The local curvature of the powerline catenary $R$ can be expressed as
\begin{equation*}R = \dfrac{T_0}{\rho g A} \cosh^2{\left\{ \dfrac{\rho g A}{T_0} \left(x - \dfrac{L}{2}\right)\right\}}.
\end{equation*}

\subsubsection{Local spatial frequency of a catenary}
One can find the frequency of the catenary curve based on the assumption that the physical properties of the powerline can be estimated and that the travel speed of is constant as it flies over the powerline. Then, the local spatial frequency of the catenary can be defined as
\begin{equation}
    \omega_s (x) = \dfrac{x}{y(x)}. 
    \label{eq:sensor_freq}   \end{equation}

\subsection{Aircraft control approach with shape adaptive actuation}

Shape adaptive actuation alters an aircraft's aerodynamics and associated stability derivatives by modifying its physical geometry. Variations induced by shape adaptiveness can be formulated as a perturbation of traditional stability derivatives \citep{Beard} by introducing a vector of generalized time-varying shape adaptive parameters $\sigma(t)$. This approach is further specialized for the purpose of this paper for camber or thickness and span shape adaptation.

This section presents the definition of the linear system and a linearization approach that incorporates adaptive dynamics of the nonlinear shape as an explicit function of $\sigma(t)$. In state-space form, a non-linear system model $\dot{x}=f(x,u,t)$ can be expressed as

\begin{equation*}
\dot{x} \approx A[\sigma(t)]x + B\left[\sigma(t)\right]u + W[\sigma(t)] \end{equation*}
with
\begin{equation*}
A[\sigma(t)] = \left. \dfrac{\partial f(x,u,t)}{\partial x}\right|_{x_0} \qquad B[\sigma(t)] = \left. \dfrac{\partial f(x,u,t)}{\partial u}\right|_{u_0},
\end{equation*}
where $W[\sigma(t)]$ represents higher-order nonlinear terms, and the state matrices vary with $\sigma(t)$, a time-varying vector that may consist of system outputs, exogenous inputs, or a combination of both. In the sections below, example dynamics of two types of shape adaptive structures for a UAV are discussed, and how they can be used as actuation to control a shape adaptive aircraft is developed.

\subsubsection{Example: Airfoil shape adaptation (Camber and Thickness)}

The introduction of laterally symmetric, shape-adaptive alterations significantly impacts the longitudinal aerodynamics, particularly the normal force and the pitch moment stability derivatives. The traditional linearized state-space dynamics model can then be written in terms of a generic morphing control input $\sigma(t)$. Consistent with traditional linearized modeling, states are defined in terms of perturbations from trim values. Shape adaptive aircraft having independent shape and flight control, as in a linear system, have been discussed in the literature \citep{camber}. An analogous linearization allows writing force and momentum as
\begin{align*}
    m \ddot{x} =& -D\cos{\alpha} + L \sin{\alpha} + T - mg\sin{\theta} \\
    m \ddot{w} =& -D\sin{\alpha} - L \cos{\alpha} + mg \cos{\theta} \\
    I_{yy} \dot{q} =& M_A - gS\sin{\theta},
\end{align*}
where the lift, drag, and moment can be described as
\begin{align*}
    L &= 0.5 \rho V^2 S C_L \\
    D &= 0.5 \rho V^2 S C_D \\
    M_A &= 0.5 \rho V^2 Sc C_m. \end{align*}
The aerodynamic coefficients directly affected by the camber or thickness shape $\mu$ can be written in a $C_{[L, D,m]}(\alpha, V,\mu)$ form similar to stability derivatives as
\begin{align*}
    C_L &= C_L(\alpha,V,\mu) = C_{L0} + C_{L\alpha}\alpha + C_{LV}\Delta V/V + C_{L\mu} \mu \\
    C_D &= C_{D0} + C_{D\alpha}\alpha + C_{D\alpha^2}\alpha^2 + C_{DV}\Delta V/V + C_{D\mu} \mu \\
    C_m &=C_{m0} + C_{m\alpha}\alpha + C_{mV}\Delta V/V + C_{m\mu} \mu.\end{align*}

\subsubsection{Example: Variable span shape adaptation}
The use of variable wingspan to improve flight performance and control authority of high endurance, medium-altitude UAV has been investigated in detail, such as \citep{span_morphing}. A linear parameter varying approach similar to \citep{LPV} is used to obtain
\begin{align*}
    \ddot{x} &= \dot{u} + qw - rv + g\sin{\theta} \\
    \ddot{y} &= \dot{v} + ru - pw + g\cos{\theta}\sin{\phi} \\
    \ddot{z} &= \dot{w} + pv - qu + g\cos{\theta}\cos{\phi} \\
    \dot{\phi} &= p + tan\theta(q\sin{\phi}+r\cos{\phi}) \\
    \dot{\theta} &= q\cos{\phi} - r\sin{\phi} \\
    \dot{\psi} &= (q \sin{\phi}+r\cos{\phi})\sec{\theta} \\
    p &= \dot{\phi} - \dot{\psi} \sin{\theta} \\
    q &= \dot{\theta}\cos{\phi} + \dot{\psi}\cos{\theta}\sin{\phi} \\
    r &= \dot{\psi}\cos{\theta}\cos{\phi} - \dot{\theta}\sin{\phi} 
\end{align*}
\begin{align*}
    \sum L &= \sum (L_0+\Delta L) = \dot{p}I_x + p \dot{I}_x - \dot{q}I_{xy} - q\dot{I}_{xy}-\dot{r}I_{xz} \\& -r\dot{I}_{xz}-qpI_{xz} - q^2I_{yz}+qrI_z-rqI_y+r^2I_{yz}\\
    \sum M &= \sum (M_0+\Delta M) = \dot{q}I_y + q \dot{I}_y - \dot{p}I_{xy} - p\dot{I}_{xy}-\dot{r}I_{yz} \\& -r\dot{I}_{yz}+rqI_{x} - rqI_{x} - r^2I_{xz}+p^2I_{xz}+pqI_{yz}-rqI_z \\
    \sum N &= \sum (N_0+\Delta N) == \dot{r}I_z + r \dot{I}_z - \dot{p}I_{xz} - p\dot{I}_{xz}-\dot{q}I_{yz} \\& -q\dot{I}_{yz}-p^2I_{xy} + pqI_{y} - prI_{yz} - qpI_x + q^2I_{xy} + qrI_{xz}.\end{align*}
Assuming the adaptation of the symmetric span shape and using $\delta y$ and $\delta \dot{y}$ as the adaptive parameters of the shape, we arrive at 
\begin{align}\sum L &= \dot{p}I_{x0}+qr(I_{z0}-I_{y0})+2m_s\dot{\delta}y(y_0p+\delta yp)\\ 
\sum M &= \dot{q}I_{y0}+rq(I_{x0}-I_{z0})\\
\sum N &= \dot{r}I_{z0}+pq(I_{y0}-I_{x0})+2m_s\dot{\delta}y(y_0r+\delta yr).\end{align}

The linearized aircraft model augmented with non-linear shape adaptation thickness and span shape $\sigma = [\mu, \delta y, \delta \dot{y}]$ can be written in terms of normalized aerodynamic stability derivatives as
\begin{equation} \label{eq:shape adaptive}
    \begin{bmatrix} \dot{u} \\ \dot{w} \\ \dot{q} \\ \dot{\theta} \\ \dot{h} \end{bmatrix} = \begin{bmatrix} X_u(\sigma) & X_w(\sigma) & X_q & -g\cos{\theta_0} & 0 \\ Z_u(\sigma) & Z_w(\sigma) & u_0 & -g\sin{\theta_0} & 0 \\ M_u(\sigma) & M_w(\sigma) & M_q(\sigma) & 0 & 0 \\ 0 & 0 & 1 & 0 & 0 \\ \cos{\theta_0} & \sin{\theta_0} & 0 & 0 & 0 \end{bmatrix} \begin{bmatrix} u \\ w \\ q \\ \theta \\ h \end{bmatrix}.
\end{equation}

\subsubsection{Structural Actuation}
Shape-adaptive aircraft must include an actuation mechanism. Servos are commonly used to modify the airfoil shape. A second-order transfer function with delay was used to represent the servo, and a slew-rate limit was also included \citep{servo_slew_rate}. %This part says nothing significant or accurate, removed! The slew rate constraint ensures that shape-adaptive airfoil movement is limited to a maximum rate. The servo-actuation transfer function with a slew-rate limit is particularly useful in shape-adaptive aircraft applications. The rate at which the shape adaptive airfoils change shape is critical to the aircraft's structural stability and efficiency. By limiting the shape-adaptive airfoil movement rate, the system can prevent overshoot and instability, ensuring that the aircraft remains stable and efficient throughout the flight. This also plays a critical role in the control design and how the aircraft trajectory behaves in applications of powerline catenary tracking.

% \begin{figure}[hbt]    \centering
%     \includegraphics[width=0.425\textwidth]{fig/morphing_act.pdf}
%     \caption{Actuator response}
%     \label{fig:actuator_dynamics}
% \end{figure}

\section{Catenary frequency mapping control design}
Flight speeds between pylons are in the range of 1-3 seconds, placing them near the Phugoid mode of a Group 1 UAV \citep{FAA}.  The Phugoid mode describes a tradeoff between airspeed and altitude, and leveraging this natural mode may be an especially relevant method of achieving the height variation needed for powerline tracking. A concise control architecture can be developed using aerodynamic shape adaptation to match the Phugoid mode's temporal frequency to the local powerline spatial frequency in Equation~(\ref{eq:sensor_freq}). The powerline's local spatial frequency prescribes the required aerodynamic shape for close tracking; equivalently, the Phugoid's spatial wavelength is modulated to match the powerline.

To implement this frequency-matching approach, Equation \eqref{eq:shape adaptive} is linearized with respect to the nonlinear shape adaptive parameter $\sigma(t)$ about $\sigma_0=0$ to obtain  
\begin{align}\dot{x}(t)&=Ax(t) +B_\sigma\sigma(t) x(t), \notag\\
 &=\left[A+B_\sigma\sigma(t)\right] x(t), \label{eq:lin_shape adaptive}\end{align}
with %\hspace{-2em}\text{with} \notag\end{align}
%\begin{bmatrix} \dot{u} \\ \dot{w} \\ \dot{q} \\ \dot{\theta} \\ \dot{h} \end{bmatrix}= \left\{ 
\begin{align*}A&=\begin{bmatrix} X_u & X_w & X_q & -g \cos\theta_0 & 0 \\ Z_u & Z_w & Z_q & -g \sin\theta_0 & 0 \\ M_u & M_w & M_q & 0 & 0 \\ 0 & 0 & 1 & 0 & 0 \\ \sin\theta & -\cos\theta & 0 & u \cos\theta+ w \sin\theta & 0 \end{bmatrix}, \notag \\
B_\sigma&=\begin{bmatrix} X_{u\sigma} & X_{w\sigma} & 0 & 0 & 0 \\ Z_{u\sigma} & Z_{w\sigma} & 0 & 0 & 0 \\ M_{u\sigma} & M_{w\sigma} & M_{q\sigma} & 0 & 0 \\ 0 & 0 & 0 & 0 & 0 \\ 0 & 0 & 0 & 0 & 0 \end{bmatrix}.%\sigma \right\} \begin{bmatrix} u \\ w \\ q \\ \theta \\ h\end{bmatrix}
\notag\end{align*}
Analyzing the long period mode from Eqn.~\eqref{eq:lin_shape adaptive}, the Phugoid (temporal) frequency is
\begin{equation}
    \label{eq:phug_freq}\omega_{t}(\sigma) = \sqrt{\dfrac{g(Z_u + Z_{u\sigma}(\alpha,\sigma)\sigma)}{Z_q}}.\end{equation}
The catenary's local spatial frequency $\omega_s$ can be used to find the required shape adaptation analytically, first as a requirement on the aircraft Phugoid frequency $\omega_t$, then via aerodynamic sensitivity as a requirement on the morphing parameter $\sigma$.  The spatial frequency matching requirement is
\begin{equation}\omega_{s}(x) =\dfrac{1}{u_0}\omega_t(\sigma),\label{eq:omegaReqt}\end{equation}
or, substituting Eqn.~\eqref{eq:sensor_freq} and \eqref{eq:phug_freq}  into \eqref{eq:omegaReqt}, 
\[\dfrac{u_0 x}{y(x)} = \sqrt{\dfrac{g(Z_u + Z_{u\sigma}(\sigma,\alpha)\sigma)}{Z_q}}.\]

\begin{figure}[!htb]
    \centering
    \includegraphics[width=0.45\textwidth]{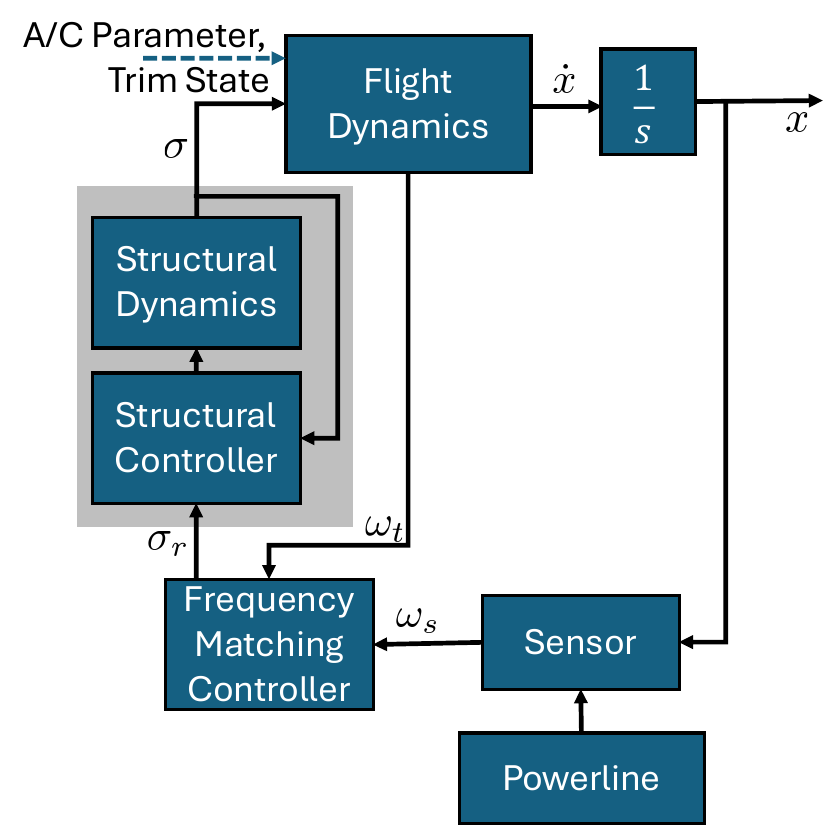}
    \caption{Controller architecture for phugoid-catenary mapping}
    \label{fig:control_flow} \end{figure}

Solving for terms dependent on morphing $\sigma$, one has
\begin{equation*}Z_{u\sigma}(\sigma,\alpha)\sigma = \dfrac{Z_q u_0^2x^2}{gy(x)^2} - Z_u. \label{e:Zusigma}\end{equation*}
which can be simplified to
\begin{equation}\label{eq:sigma}
    Z_{u\sigma}(\sigma,\alpha)\sigma - \dfrac{Z_q u_0^2x^2}{gy(x)^2} + Z_u = 0.\end{equation}
Using the lookup table for a given set of the angle of attack $\alpha$ and the aerodynamic parameters $Z_{u\sigma}$, Eqn. \eqref{eq:sigma} can be solved using numerical solvers such as Newton Raphson method to extract an adaptive parameter of the analytical shape as a function of the distance traveled from the post, provided that there is post-inflection point. The resulting feedforward controller architecture is shown in Fig.~\ref{fig:control_flow}.

\subsection{Design Simulations}
To generalize airframe selection and the range of shape adaptive structures needed, using the definition of mission space and the phugoid-catenary matching controller for a generic reference aircraft example chosen through trial and error in simulation that gives adequate results and using that as a base, a trend search is performed using unconstrained nonlinear least squares-based optimization to get us aerodynamic parameters required for a given powerline structure.

The optimization performed is
\[\hat{\sigma}(t) \in \text{argmin}_\sigma \sum_{i=1}^{i=N}[s_\text{ref} - s\left(x(t),\sigma(t_i)\right)]^2,\]
where $s_\text{ref}$ is the reference powerline trajectory (here assumed known at a given time and location) and $s(x(t),\sigma(t_i))$ is the actual trajectory for the aircraft for a given time series of morphing $\sigma(t_i)$ and the aircraft state $x(t)$.

\subsubsection{Simulation Case Study: Generic aircraft (Group I type)} %>> change the aircraft

\paragraph{Shape Adaptive Aircraft Design}
This study simulated a reference aircraft that includes variations in airfoil camber and airfoil thickness. A nominal NACA2412 airfoil capable of adapting to the shape from NACA1412 to NACA4416 for the maximum camber location change and from NACA2406 to NACA2416 for thickness change.

\begin{figure}
    \centering
    \includegraphics[width=0.45\textwidth]{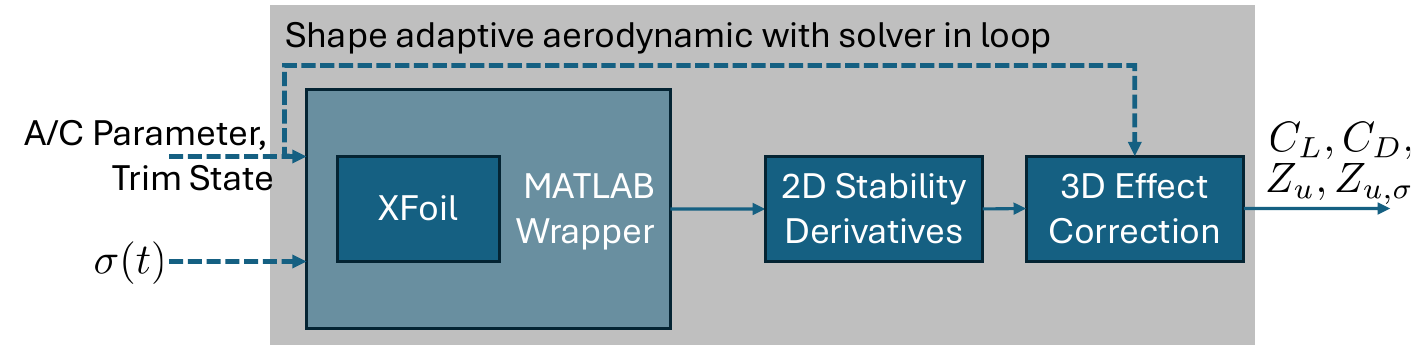}
    \caption{Shape adaptive solver in loop method of aerodynamic parameter calculation}\label{fig:SASIL}
\end{figure}
\begin{figure}
    \centering
    \includegraphics[width=0.45\textwidth]{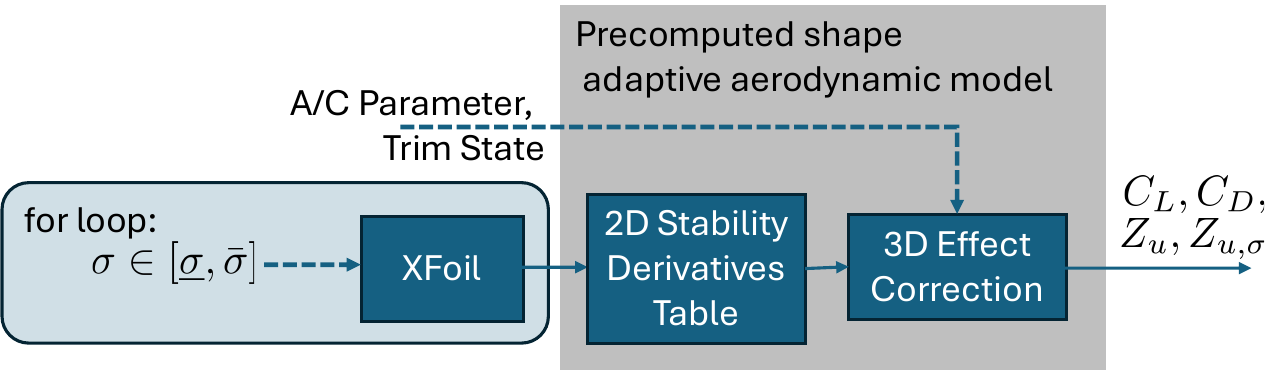}
    \caption{Pre computed shape adaptive aerodynamic parameter calculation}\label{fig:PCSA}
\end{figure}

To simulate continuous shape adaptive aerodynamics, XFoil and Matlab were used to characterize the aerodynamic parameters for the aircraft flight envelope using two methods as illustrated in Fig.~\ref{fig:SASIL} and Fig.~\ref{fig:PCSA}, where the stability derivatives related to shape adaptive parameters are calculated in loop and pre-calculated using prior knowledge of shape adaption limits ($\underline{\sigma},\bar{\sigma}$) where $\underline{\sigma}$ is the lower limit of morphing and $\bar{\sigma}$ is the upper limit of the morphing parameter. 

The method results in aerodynamic stability derivatives as functions of shape adaptive parameters. A example of Eqn.~\eqref{eq:lin_shape adaptive} maneuvering about trim is
\begin{equation}\begin{split}\begin{bmatrix} \dot{u} \\ \dot{w} \\ \dot{q} \\ \dot{\theta} \\ \dot{h} \end{bmatrix} = \left(
\begin{bmatrix} -0.074 & -0.122 & 0 & -9.81 1 & 0 \\ -1.535 & -7.457 & 25 & 0 & 0 \\ 2.689 & -5.850 & -32.945 & 0 & 0 \\ 0 & 0 & 1 & 0 & 0 \\ 1 & 0 & 0 & 0  & 0 \end{bmatrix}\right. + \\ 
  \left.\begin{bmatrix}-0.332 & -2.096 & 0 & 0 & 0 \\ -3.960 & -15.336 & 0 & 0 & 0 \\ 25.329 & 95.192 & -160.32 & 0 & 0 \\ 0 & 0 & 0 & 0 & 0 \\ 0 & 0 & 0 & 0 & 0 \end{bmatrix}\sigma\right)\begin{bmatrix} u \\ w \\ q \\ \theta \\ h \end{bmatrix}.\end{split}\label{e:trim}\end{equation}

\paragraph{Simulation Method}
To illustrate the proposed controller operation, this approach is implemented in MATLAB, where a high-fidelity nonlinear shape adaptive aircraft is simulated with the control input developed using a linearized shape adaptive aircraft dynamics representation as developed above. The simulation includes a reference power line with 30m tall towers spaced at 70m intervals.

The controller has local information about the powerline geometry to estimate the catenary spatial frequency and full state information about the aircraft from the dynamics to predict the Phugoid temporal frequency for phugoid-catenary mapping; the controller is re-initialized when crossing each tower.

\subsubsection{Numerical parameter space exploration} % >>Change the name of this section
To optimize the design of the powerline unmanned surfer (PLUS) for diverse operational environments, simulations to explore the interplay between different aircraft design parameters and various powerline structures were conducted. This step was instrumental in identifying the most effective combinations of aircraft wingspan and chords for powerline tracking performance across a range of powerline geometries.

The trend search involved running a series of simulations, each iterating over a set of predefined span and chord values for the UAV. These values were tested against multiple powerline structures, characterized by different catenary curve parameters such as sag, tension, and span length. The primary objective was to determine how changes in the UAV's physical dimensions influenced its ability to adapt and maintain efficient tracking of the powerlines, especially in varying environmental conditions.

\paragraph{Aircraft parameters}
The simulations systematically varied two UAV parameters: chord length and wingspan. The wing's chord (leading edge to trailing edge distance) was varied, exploring the effects of change in chord length for group I type UAVs. The chord lengths varied from $0.203\:\text{m}$ to $0.305\:\text{m}$ with increment of $0.51\:\text{m}$.  The wingspan (distance between the tips of the wings) was altered in each simulation run, testing a range of typical Group I UAVs. Wingspans ranged from $1\:\text{m}$ to $1.8\:\text{m}$ with increments of $0.4\:\text{m}$ 

\paragraph{Environmental parameters}
The variations were evaluated across a spectrum of powerline geometries ranging from low voltage to ultra high voltage (>800kV), as outlined in Table~\ref{tab:powerline}, to quantify the spacing and height dimensions of the pylons. Furthermore, the impact of environmental factors on the powerlines was standardized by the percentage of sag throughout the pylon span.
\begin{table*}[!hbtp]\centering   \begin{tabular}{p{4.1cm}rrr}  \toprule %was p{2.1cm} for two column format
        \textbf{Label} & \textbf{Voltage, kV} & \multicolumn{2}{c}{\textbf{Pylon geometry}}\\\cmidrule(lr){1-1}\cmidrule(lr){2-2}\cmidrule(lr){3-4}
        & & Height, m &Spacing, m \\
        \midrule
        Low voltage (LV) & Up to 1 & 10 - 15 & 30 - 50 \\
        Medium voltage (MV) & 1 - 69 & 15 - 30 & 50 - 150 \\
        High voltage (HV) & 69 - 230 & 30 - 50 & 150 - 400 \\
        Extra high voltage (EHV) & 230 - 500 & 50 - 80 & 300 - 500 \\
        Ultra high voltage (UHV) & 800+ & 80+ & 500+ \\        \bottomrule    \end{tabular}
    \caption{Powerline class and structural parameters \citep{EARI_data}.}  \label{tab:powerline}
\end{table*} 
This approach provided a mechanism to quantify the UAV's performance across varying environmental geometry.

\section{Results and discussion}
\subsection{Thickness morphing experiment}
\begin{figure}[hbt]    \centering
    \includegraphics[width=0.4\textwidth]{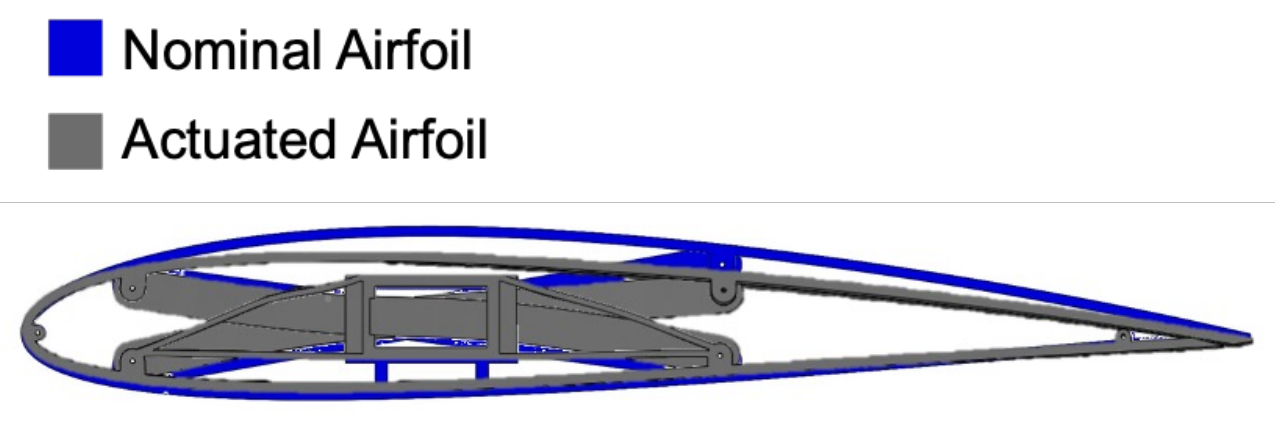}
    \includegraphics[width=0.4\textwidth]{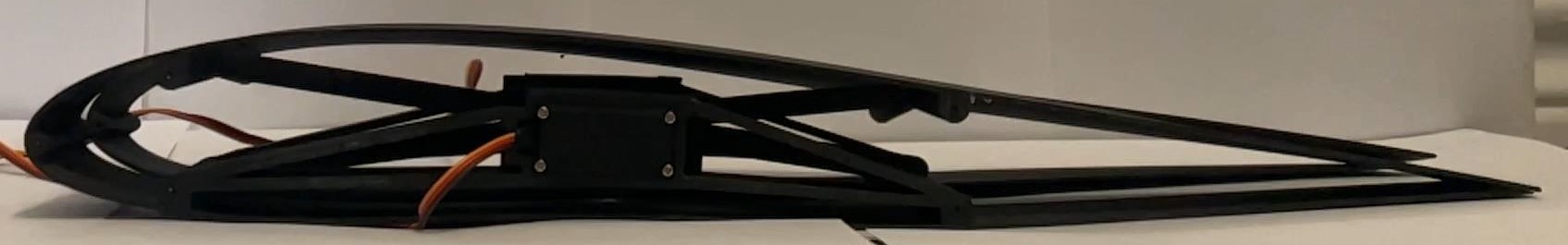}
    \caption{Thickness morphing airfoil actuator used in experimental identification}\label{fig:sysid_setup}\end{figure}

A representative servo actuation system was manufactured and the identification of the time domain system was used to determine the transfer function from the experimental data. The experiment used the additively manufactured airfoil shown in Fig.~\ref{fig:sysid_setup} which has a degree of freedom of thickness actuated by the servo.  A camera operating at 120 frames per second was used to digitize the thickness variation using the MATLAB image processing toolbox \citep{matlab_image}. The actuator system received a multistep input command and the recorded input/output signals were used to quantify the response time, the maximum deflection angles, and the effects of the slew rate limit of $0.11 s/60^\circ$ degree rotation.

\begin{figure}[!hbtp]\centering
    \includegraphics[width=0.45\textwidth]{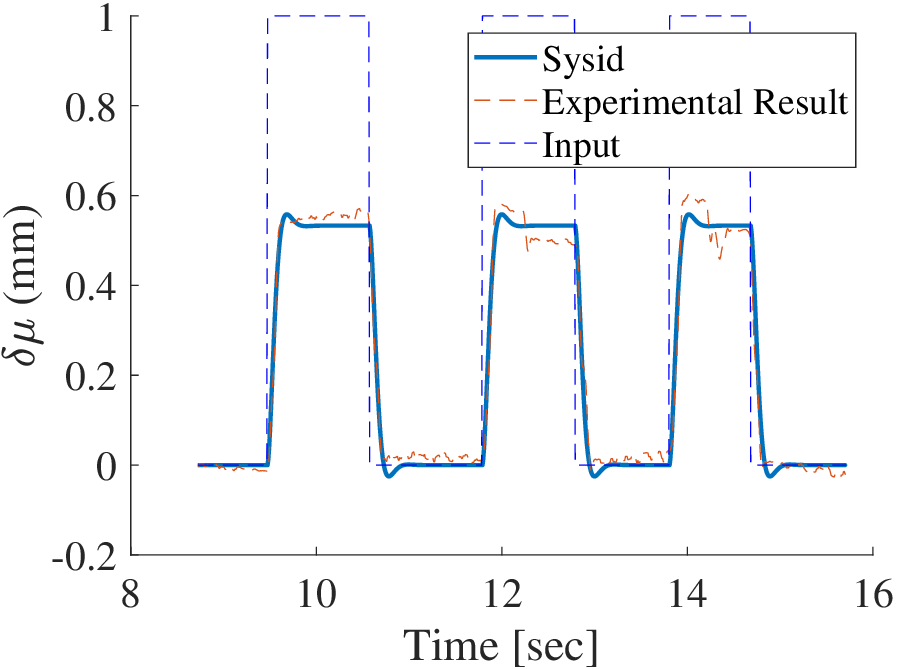}
    \caption{Morphing data fit to a second order system with time delay}\label{fig:sysid_resutls}\end{figure}

A multiple-step stimulus of $3^\circ$ servo deflection was used for identification, and the identification of the time-domain system considered three transfer function model structures: first order, second order system, and second order with time delay. The identification results in Table \ref{tab:sysid_tab} indicate that a 2nd order model including transport delay best describes the response.
\begin{table}[!hbtp]    \centering
    \begin{tabular}{rccc}\toprule
        \multicolumn{1}{c}{\textbf{Parameter}} & \multicolumn{3}{c}{\textbf{Model structure}}\\ \cmidrule(lr){1-1} \cmidrule(lr){2-4}
        %\cmidrule{2-4}& 1\textsuperscript{st} order& 2\textsuperscript{nd} order & 2\textsuperscript{nd} order and delay \\
        &\multicolumn{1}{c}{1\textsuperscript{st} order} &\multicolumn{2}{c}{2\textsuperscript{nd} order}\\
        \cmidrule(lr){2-2}\cmidrule(lr){3-4}
        & & Nominal&With delay\\                  
        \midrule
        Gain & 0.01333 & 0.01333 & 0.01333 \\
        Time constant & 1.0s & - & - \\
        Damping ratio & - & 0.919 & 0.45 \\
        Natural frequency & - & 1.0Hz & 1.0Hz \\
        Time delay & - & - & 0.05s \\
        Fit accuracy & 86\% & 91\% & 93\% \\
        \bottomrule
    \end{tabular}
    \caption{Structural actuation system identification results}    \label{tab:sysid_tab}\end{table} \\
The identified thickness morphing transfer function was 
\begin{equation}
    Y(s) = \dfrac{0.01333}{s^2-1.838s+1}\left(1 - e^{-0.005s}\right),\label{e:morph_tf}\end{equation}
as seen in Fig.~\ref{fig:sysid_resutls}.

%The resulting transfer function in Eqn.~\eqref{e:morph_tf}, derived from these experimental observations, not only validated our theoretical model but also offered refinements. 

\subsection{Simulation results}
\begin{figure}[!hbtp]    \centering
    \includegraphics[width=0.45\textwidth]{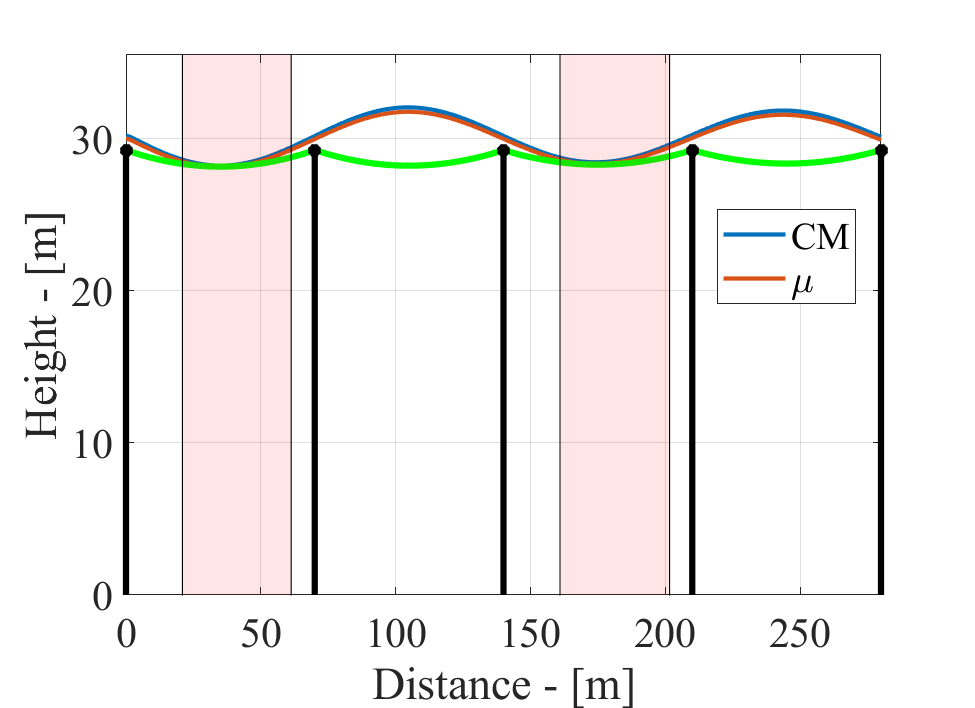}
    \caption{Powerline tracking for shape adaptive aircraft}
    \label{fig:airtrej}
\end{figure}

\begin{figure}[!hbtp]    \centering
    \includegraphics[width=0.45\textwidth]{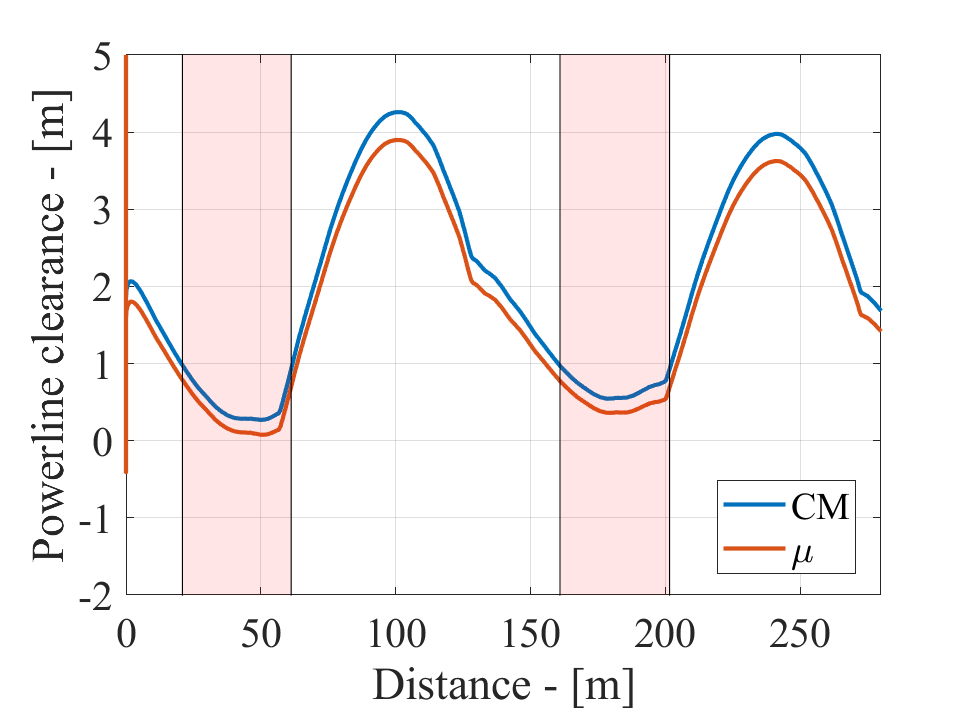}
    \caption{Error between aircraft trajectory and powerline}
    \label{fig:errtrej}
\end{figure}

\begin{figure}[!hbtp]    \centering
    \includegraphics[width=0.45\textwidth]{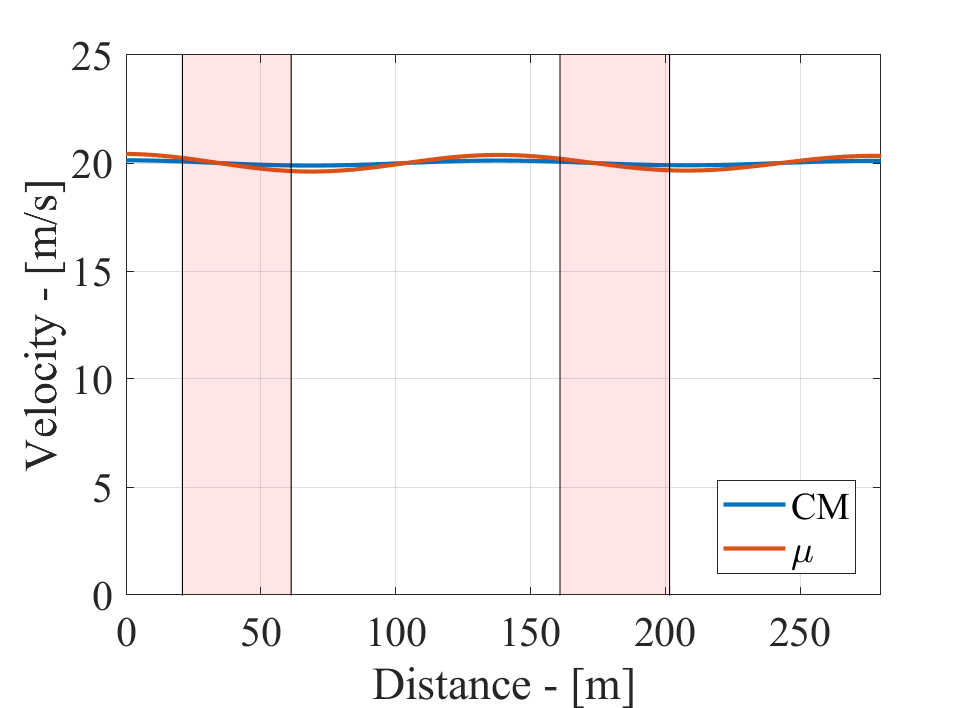}
    \caption{Velocity deviation for the powerline tracking}
    \label{fig:veltrej}
\end{figure}

\begin{figure}[!hbtp]    \centering
    \includegraphics[width=0.45\textwidth]{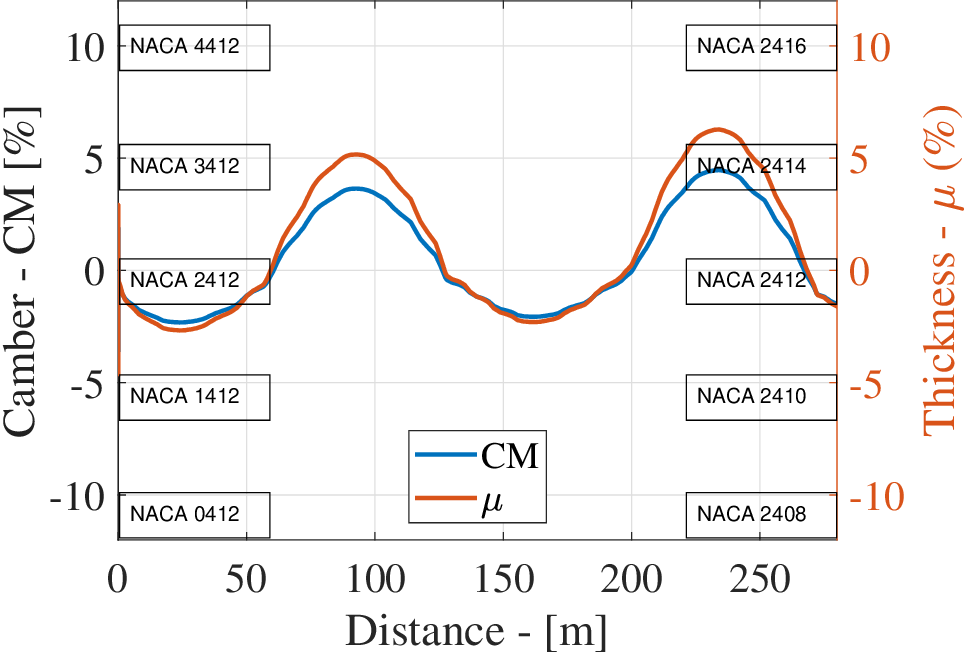}
    \caption{Continuous shape adaptive input to the system for powerline tracking}
    \label{fig:morphing}
\end{figure}

The results of the aircraft trajectory based on the implemented controller are shown in Fig.~\ref{fig:airtrej}. The tracking of the power line is achieved periodically in alternating troughs, leading to low and high levels of cyclic clearance, as shown by the shaded regions in Fig.~\ref{fig:airtrej}. The powerline clearance shown in Fig.~\ref{fig:errtrej} shows that for the chosen aircraft the variation of - 3\% to 6.3\% of the camber or the variation of -3.2\% to 5.5\% thickness shown in Fig.~\ref{fig:morphing} achieves <1m powerline spacing for 88.7m of the 260m span, or 34\% of the longitudinal region. Fig.~\ref{fig:veltrej} shows the velocity deviation during the controller implementation. 

\subsection{Parameter space exploration}
The parameter space exploration examined the sensitivity to wingspan, powerline span, and wing chord. The proficiency requirement is quantified using the maximum aerodynamic change required by a morphing that is found by considering the unmorphed aerodynamic coefficient relative to the maximal morphing seen in the trial.  For example, for the case of lift coefficient $C_L$, the maximal $C_L$ variation observed is found by parameterising lift coefficient as a function of morphing, eg $C_L(\alpha,\sigma)$.  Then, we find the maximum morphing parameter value and the time at which it occurs, eg, defining
\begin{align}
t_m&=\arg \max_t{\sigma(t)}, \quad \forall t\in[0,t_f]\\
\bar{\sigma}&=\max_t{\sigma(t)},   \quad \forall t\in[0,t_f]\end{align}
for a trial lasting from time $t=0$ to final time $t_f$, then the maximum change in lift coefficient demanded by morphing in that trial is 
\[\Delta C_{L,m} = C_L\left(\alpha(t_m), \bar{\sigma}\right) - C_L\left(\alpha_0,0\right),\]
where $\alpha_0$ denotes the trimmed angle of attack.

%Combined the above 3 equations:\[\Delta C_{L,m} = C_L\left(\alpha(\argmax{\sigma(t)}), \max_t{\sigma (t)}\right) - C_L\left(\alpha_0,\sigma=0\right) \quad \forall t\in[0,N],\] where $\alpha_0$ denotes the trimmed angle of attack.

% This term is bounded by 1 based on
% \[\max {\Delta C_{L,m}} \leq \max{(C_L\left(\alpha(t_m), \bar{\sigma}\right)} - \min{C_L\left(\alpha_0,0\right)}.\]

$\Delta C_{L,m}$ was computed for a total of 4,500 cases, covering 25 trials for each combination of 4 pylon span distances that cover LV to EHV powerlines,  3 airfoil chords, 3 aircraft wingspans and 5 sag level combinations.

\begin{figure}[!hbtp]    \centering
    \includegraphics[width=0.45\textwidth]{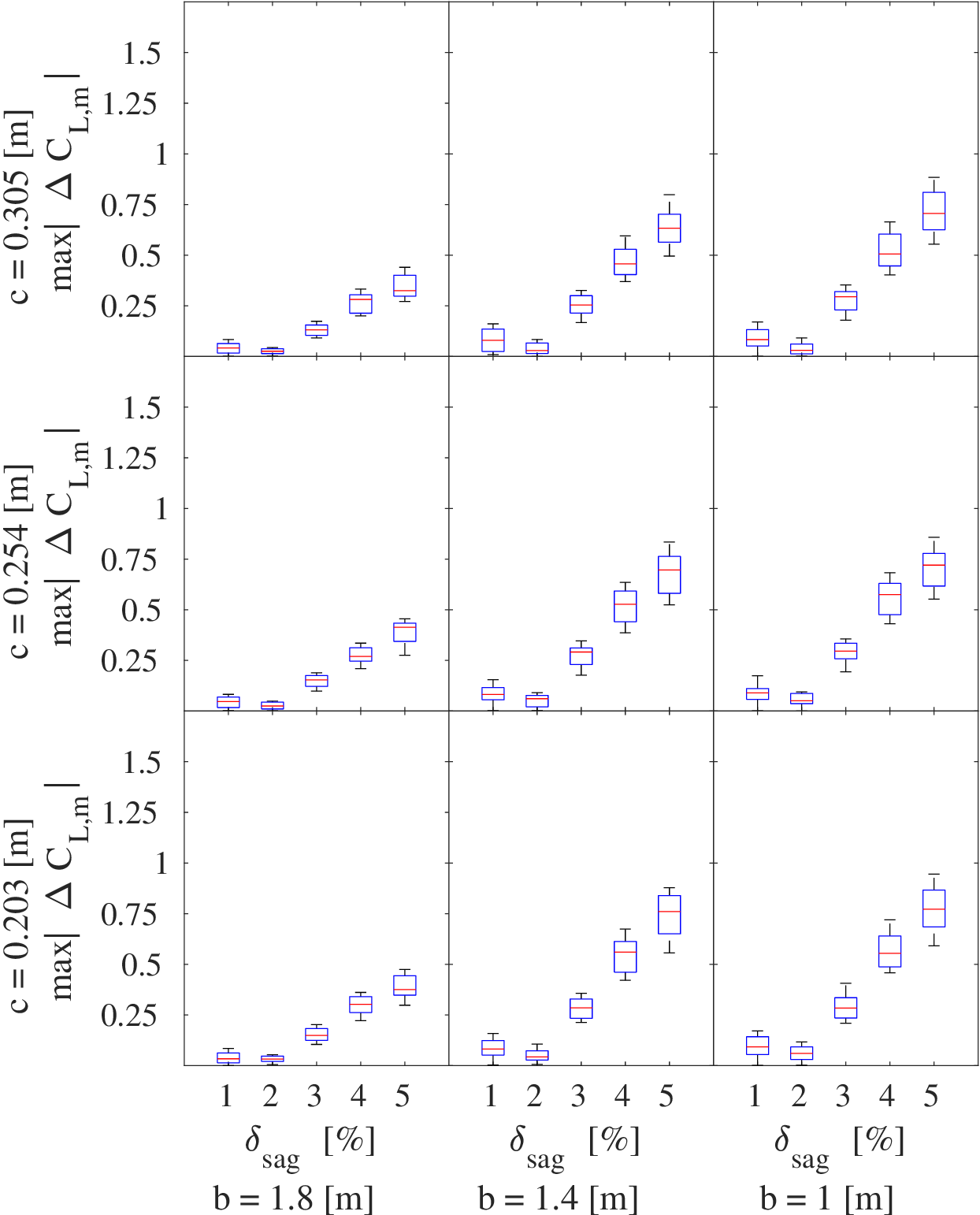}\\
    \caption{Parameter space exploration results for LV powerlines (40m pylon span)}
    \label{fig:trendsearch_d40}
\end{figure}
\begin{figure}[!hbtp]    \centering
    \includegraphics[width=0.45\textwidth]{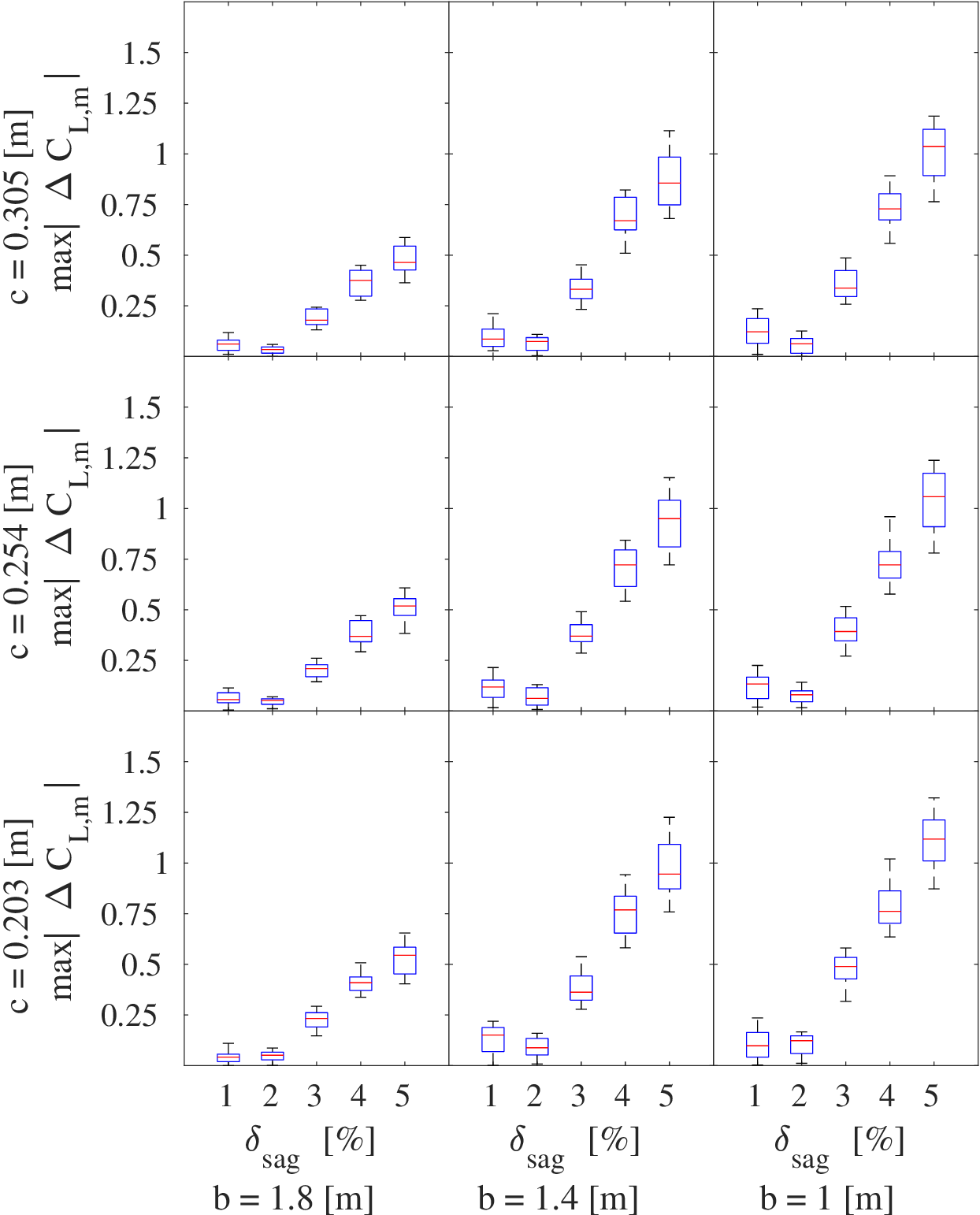}\\
    \caption{Parameter space exploration results for MV powerlines (100m pylon span)}
    \label{fig:trendsearch_d100}\end{figure}
\begin{figure}[!hbtp]    \centering
    \includegraphics[width=0.45\textwidth]{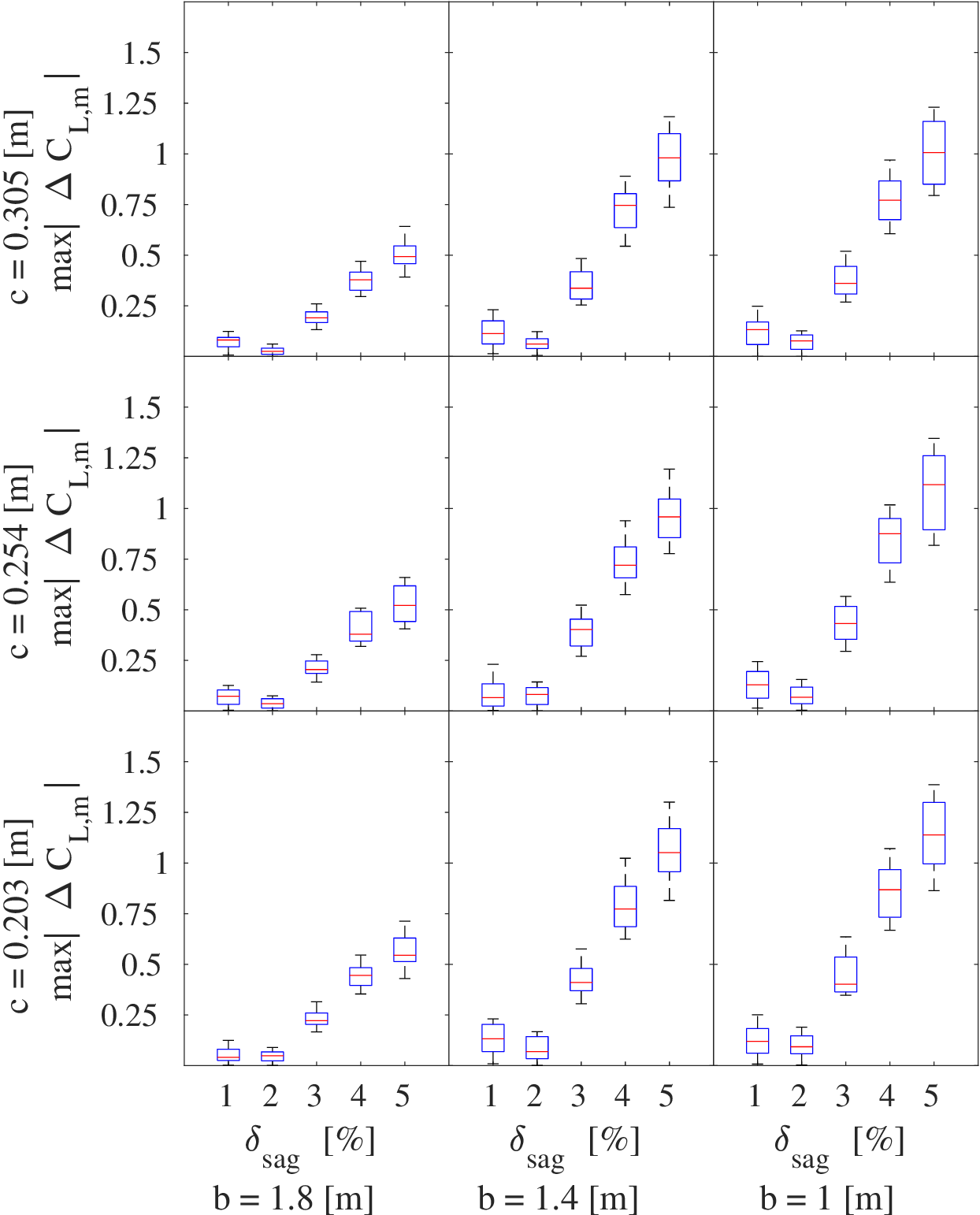}
    \caption{Parameter space exploration results for HV powerlines (300m pylon span)}
    \label{fig:trendsearch_d300}\end{figure}

\begin{figure}[!hbtp]    \centering
    \includegraphics[width=0.45\textwidth]{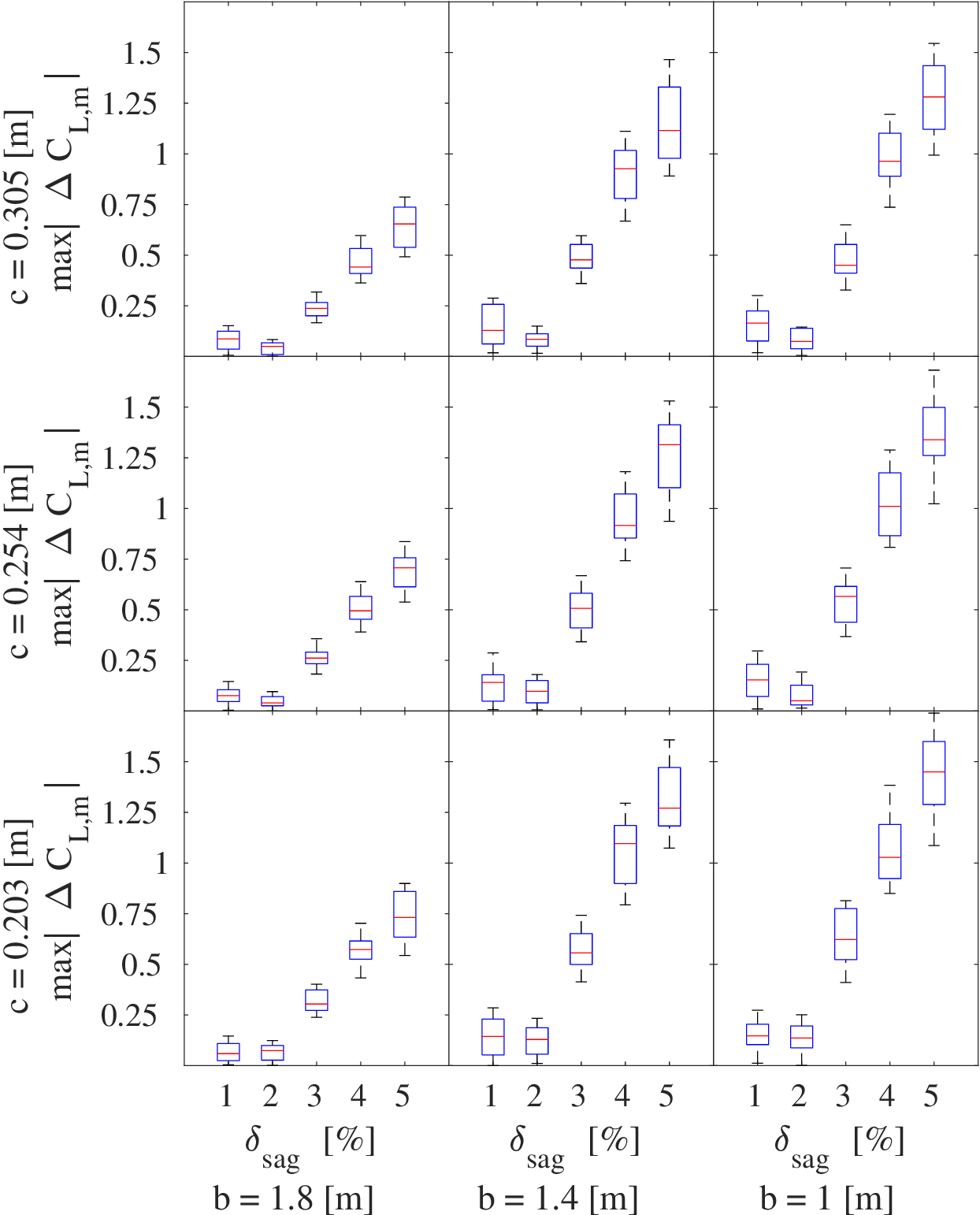}
    \caption{Parameter space exploration results for EHV powerlines (500m pylon span)}
    \label{fig:trendsearch_d500}
\end{figure}
Mean morphing-driven changes in $C_L$ (as quantified by $\max{|\Delta C_{L,m}}|$) increases with decreases in either  mean aerodynamic chord (MAC) decreases or wing area, with the area effect is most visible in Fig.~\ref{fig:trendsearch_d500}.  Both are likely related to the need for higher lift coefficients on smaller aerodynamic surfaces.  

\begin{figure*}[!htbp]
    \centering
    \includegraphics[width=0.6\textwidth]{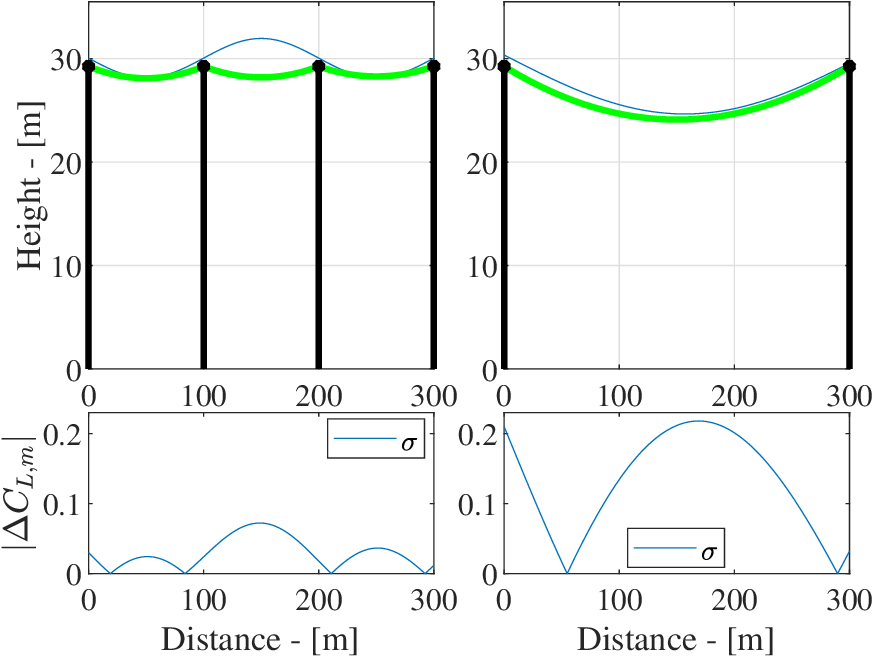}
    \caption{Illustration of lowered aerodynamic actuation needed for powerline wavelengths being near an integer factor of the aircraft's (c=0.254m, b=1.4m) Phugoid wavelength}
    \label{fig:nearintegerillustration}
\end{figure*}

Across Figs.~\ref{fig:trendsearch_d40}-\ref{fig:trendsearch_d500}, the 2\% powerline sag results consistently show lower $\max{\Delta C_{L,m}}$ requirements. This lower actuation requirement is likely due to the 2\% sag powerline wavelengths being near an integer factor of the aircraft's Phugoid wavelength. This effect is illustrated for the cases $d=100$ m and $d=300$ m in Fig.~\ref{fig:nearintegerillustration}. This lowered aerodynamic actuation when Phugoid and aircraft wavelength is matched lends support for the frequency-matching approach to control, and Section \ref{ss:wavelengthPlots} analyzes this relationship.
%\footnote{Variance scales with the mean magnitude.}

%{\color{blue} VARIATION as statistical variance or variation as in observed deterministic from simulation CLm bounds?  This section talks about both mean and variation and this is liable to be confusing to the reader. \begin{description}\item[Wing span variation] Larger wingspans increase the mean $\Delta C_{L,m}$.  $\Delta C_{L,m}$ variation increases with increasing powerline sag. \item[Wing chord variation] Larger wing chords increase mean $\Delta C_{L,m}$.  The variance of $\Delta C_{L,m}$ increases as the powerline sag increases. For lower powerline sag percentages, $\Delta C_{L,m}$ remains in a similar range of values with increase in chord length. \item[Powerline sag variation] Powerline sag is the primary temperature-dependent environmental effect, and increasing sag increases $\Delta C_{L,m}$. 
%\item[Change in PL span] The $\Delta C_{L,max}$ needed to track powerline increases as the powerline span increases up to spans of length 300m.  Beyond 300m span, $\Delta C_{L,m}$ required for powerline tracking is constant, suggesting the existence of an asymptote beyond which powerlines can be tracked without shape adaptation. These higher voltage powerlines may allow harvesting  energy from farther distances, suggesting that high voltage line tracking for energy harvesting may also be achievable on traditional UAV platforms without shape adaptive structures.\end{description}}

\subsection{Phugoid wavelength variation}\label{ss:wavelengthPlots}
\begin{figure}[!hbtp]    \centering
    \includegraphics[width=0.5\textwidth]{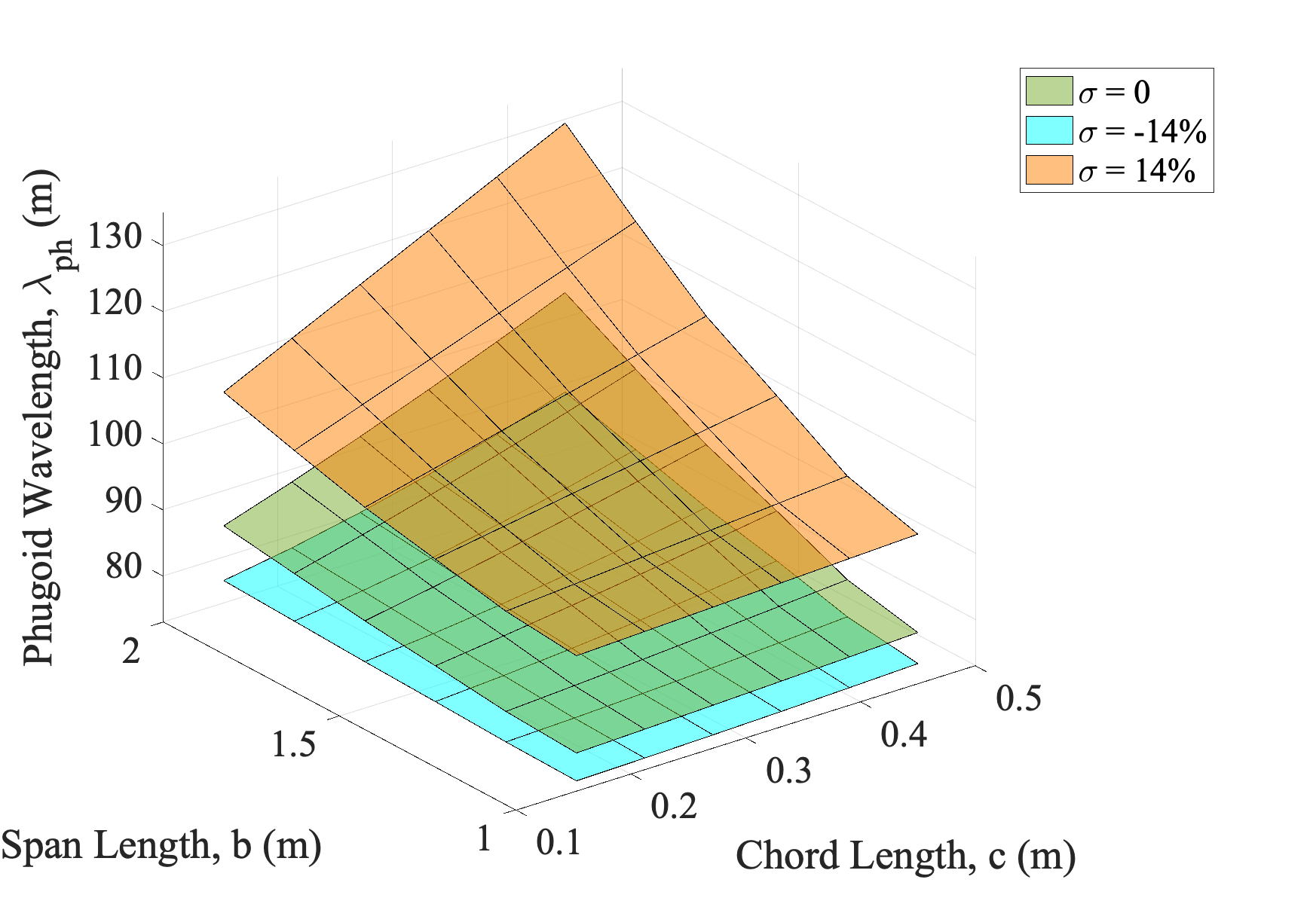}
    \caption{Phugoid wavelength during thickness morphing ($\sigma=\pm14\%$ thickness) shows the authority of thickness morphing over Phugoid length relative to span $b$ and chord $c$ variation.}    \label{fig:SurfPlot}\end{figure}

 Figure \ref{fig:SurfPlot} shows the Phugoid wavelength $\lambda_\text{ph}$ as a function of wing span, chord, and morphed thickness calculated using the Eqn.~\eqref{eq:wavelength},
 \begin{equation} \label{eq:wavelength}
     \lambda_\text{ph} = \dfrac{u_0}{\omega_t(\sigma,b,c)} \end{equation}
where, $\omega_t(\sigma,b,c)$ is the temporal frequency of a aircraft with wing span $b$ and chord length $c$ at a given morphed thickness of $\sigma$ flying at a trim speed of $u_0$.
 
Fig.~\ref{fig:SurfPlot} indicates that thickness morphing significantly affects Phugoid wavelength, and is a viable control term for matching aircraft modes to powerline spans.  Combined with Table \ref{tab:powerline}, Fig.~\ref{fig:SurfPlot} can be used to identify the feasible powerline/airframe pairings for close-range tracking and to estimate the required thickness morphing range.

%{\color{red} With more data points we might be able to establish a relationship ship of $\lambda_\text{ph}$ with AR}
\begin{figure}[!hbtp]
    \centering
    \includegraphics[width=0.5\textwidth]{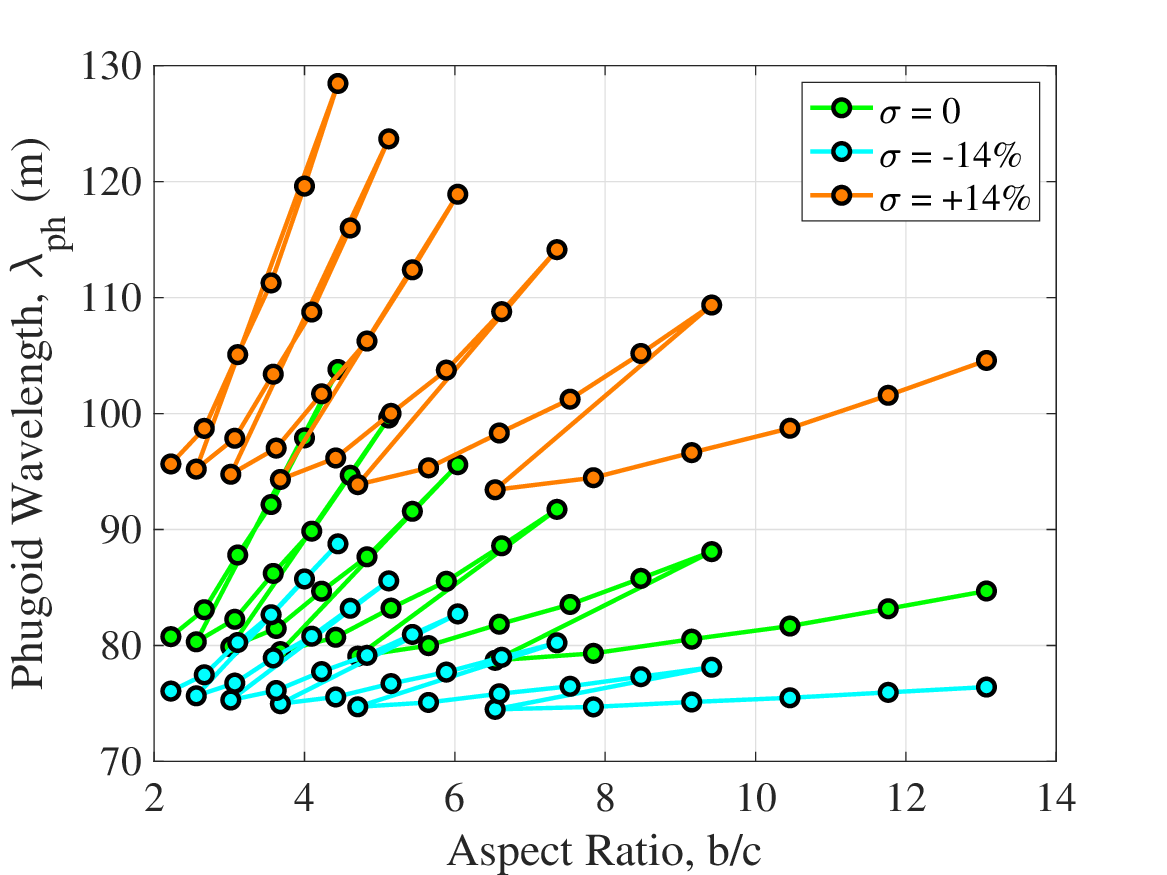}
    \caption{Phugoid wavelength during thickness morphing ($\sigma=\pm14\%$ thickness) shows the authority of thickness morphing over Phugoid length relative to aspect ratio}
    \label{fig:ARv1}\end{figure}

Figure \ref{fig:ARv1} shows that an increase in thickness provides more authority over the Phugoid wavelenth than a comparable decrease in thickness. This increased effectiveness is likely related to the higher aerodynamic sensitivity of thicker airfoils through increased lift and drag.  Support for the increase in aerodynamic sensitivity origin is also seen in that thickness morphing has higher authority over Phugoid wavelength when both spans and chord are high, as visible in Fig.~\ref{fig:SurfPlot}.  However, thickness authority is not simply a function of wing area--Fig.~\ref{fig:ARv1} also shows reduced authority at higher aspect ratios.

\subsection{Limitations and assumptions}
The open loop (feedforward) structure of this control approach does not yet account for parameter uncertainty or measurement noise, and incorporating a sensor-based feedback controller with a dynamic estimator could yield improvements.  The approach leverages actuation through a generalized aerodynamic change; thus, camber or thickness and span shape is a design choice. The finding that these modes are feasible does not indicate that other actuation modes are infeasible; rather, alternate paths may achieve this variation, including traditional empennage actuation for spans greater than 300m. Similarly, while higher-fidelity wind tunnel and experimental aerodynamic studies may improve the accuracy of aerodynamic parameter variation relative to panel methods, the frequency matching framework introduced here remains feasible when the configuration change is able to modulate aerodynamic force magnitude \citep{windunnelvsXfoil,GALFFY2019FixedWingPathFollowing}.

\section{Conclusion}
This study presents a concise dynamic model and simulation of a shape-adaptive aircraft capable of camber thickness and span shape designed for a challenging longitudinal powerline tracking mission, such as inflight recharging and perching to the powerline for extension of the flight range. The aerodynamic parameters of the aircraft are directly dependent on the time-varying aerodynamic forces and moments, which are a function of the change in the shape of the wing by the shape adaptive command ($\mu$-camber or thickness adaptive parameter). The coupled longitudinal dynamic equation of the shape adaptive process is derived by simplifying the longitudinal dynamic responses of the wing shape adaptive process. Quasi-steady aerodynamic assumptions are used to numerically simulate the dynamics.

A novel approach in this study is the implementation of a frequency-matching (or wavelength-matching) technique that dynamically adjusts the local temporal frequencies of the aircraft's natural modes to match the spatial frequency of the powerline's catenary curve. This was achieved through the generalized linearization of the adaptive parameters of the shape. The results demonstrate the potential for a control strategy capable of achieving alternating periods of low clearance tracking and antiphase oscillation. Furthermore, we integrated an open-loop control strategy that manipulates camber or thickness to maintain less than $1$ meters of powerline clearance for more than 34\% of the line, despite the challenges of alternating coverage. This dynamics and controller development lays a strong foundation for low clearance powerline tracking. Additionally, our findings suggest that for group I Unmanned Aerial Vehicles (UAVs) operating at EHV (Extra High Voltage) levels and above, intricate shape adaptation may not be required in order to fly low clearance energy harvesting trajectories. This insight could simplify the design and operation of UAVs for specific applications, reducing the complexity and enhancing the reliability of energy harvesting missions in these high voltage environments.

\bibliography{reference}
\end{document}